\documentclass[aps,amsmath,amssymb,superscriptaddress,showpacs,nofootinbib,10pt]{revtex4-1}
\usepackage{amsmath,amssymb}
\usepackage{slashed}
\usepackage{graphicx}
\usepackage{subfigure}
\usepackage{url}
\newcommand{\beq}{\begin{equation}}
\newcommand{\eeq}{\end{equation}}

\newcommand{\be}{\begin{eqnarray}}
\newcommand{\ee}{\end{eqnarray}}
\newcommand{\mytilde}{\raise.17ex\hbox{$\scriptstyle\mathtt{\sim}$}}
\long\def\hidestart#1\hideend{}
\begin{document}
\title{Correlation and localization properties of topological charge density 
and the pseudoscalar glueball mass in SU(3) lattice Yang-Mills theory}
\author{Abhishek Chowdhury}
\author{A. Harindranath}
\affiliation{Theory Division, Saha Institute of Nuclear Physics \\
 1/AF Bidhan Nagar, Kolkata 700064, India}
\author{Jyotirmoy Maiti}
\affiliation{Department of Physics, Barasat Government College,\\
10 KNC Road, Barasat, Kolkata 700124, India}
\email{abhishek.chowdhury@saha.ac.in}
\email{a.harindranath@saha.ac.in}
\email{jyotirmoy.maiti@gmail.com}
\date{March 11, 2015}
\begin{abstract}{
Towards the goal of extracting the continuum properties,
we have studied the Topological Charge Density 
Correlator (TCDC) and the Inverse Participation Ratio (IPR) for the
topological charge density ($q(x)$) in SU(3) Lattice Yang-Mills theory
for relatively small lattice spacings including some 
smaller than those explored before.
With the help of recently proposed open boundary condition, it is possible to compute
observables at a smaller lattice spacing since {\em trapping problem} is absent.
On the other hand, the reference energy scale provided by Wilson flow allows us to study 
their scaling behavior in contrast to previously proposed smearing techniques.
The behavior of TCDC
for different lattice spacings at a fixed HYP smearing level
shows apparent scaling violations.
In contrast, at a particular Wilson flow time $t$ for all the lattice spacings investigated
(except the largest one), the TCDC data show universal behavior within our
statistical uncertainties. 
The continuum properties of TCDC are studied by investigating 
the small flow time behavior.
We have also extracted the
pseudoscalar glueball mass from TCDC, which appears to be insensitive to 
the lattice spacings (0.0345 fm $\leq a\leq$ 0.0667 fm) and agrees with
the value extracted using anisotropic lattices, within statistical errors.
Further, we have studied the localization property of $q(x)$
through IPR whose continuum behavior can be probed through the small values
of Wilson flow time and observed the decrease of IPR with decreasing Wilson flow time.
A detailed study of $q(x)$ under Wilson flow time revealed that as
Wilson flow time decreases, the proximity of the regions of positive and
negative charge densities of large magnitudes increases, and the charge
density appears to be more delocalized resulting in the observed behavior of IPR.
}
\end{abstract}
\pacs{11.15.-q, 11.15.Ha, 12.38.-t, 12.38.Gc, 12.39.Mk}

\maketitle

\section{Introduction and Motivation}

The negativity of the Topological Charge Density Correlator (TCDC)  for
non-zero distances as a consequence of the reflection positivity and the
pseudoscalar nature of the relevant local operator in Euclidean field theory
is well-known \cite{es,seiler}. The non-trivial implication of the negativity 
of TCDC for the structure of topological charge density in QCD vacuum has been 
investigated in detail \cite{ih}. Various aspects of TCDC in quenched and 
full QCD have been carried out in Refs. \cite{Hasenfratz:1999ng, ih2, 
Ilgenfritz:2007xu, Ilgenfritz:2008ia, Moran, bazavov, fb, topocorr, Bruno, Fukaya}. 
The four volume integral of TCDC gives the topological susceptibility $\chi$. 
A thorough investigation of the mechanisms leading to the suppression 
of the topological susceptibility with decreasing quark mass, based on 
the properties of TCDC has been carried out \cite{topocorr} in two-flavour QCD.
The famous Witten-Veneziano formula \cite{witten,veneziano} relates $\eta'$ 
mass to the topological susceptibility in pure Yang-Mills theory. 

For the calculation of the topological charge density ($q(x)$), various
definitions have been used in the literature. Ref. \cite{durr} uses the 
algebraic definition of the field strength tensor.  
To overcome the potential lattice artifacts associated with the algebraic 
definition of the topological charge density $q(x)$ on the lattice and  
severe singularities present in TCDC in the continuum theory, various
proposals have been studied in the literature. In Ref. \cite{del},
$q(x)$ based on Ginsparg-Wilson fermion has been employed, whereas Refs.
\cite{lus2004,gl} utilize a proposal designed to overcome short distance
singularities. A spectral projection formula designed to be free from
singularity is employed in Ref. \cite{lspb} which compares the result for
topological susceptibility $\chi$ using algebraic definition. Since 
Ref. \cite{lspb} has established that the results for $\chi$ using 
various approaches are in agreement with each other within 
statistical uncertainties, in this work 
we employ the algebraic (clover) definition for $q(x)$ unless otherwise
stated.

There remain several open issues related to the topological charge density 
$q(x)$ and TCDC in non-abelian gauge theories. An important issue is the 
scaling of TCDC as one approaches the continuum limit. It is however
well-known that current lattice gauge theory simulations employing periodic 
boundary condition in the temporal direction are handicapped by the trapping
of topological charge in a particular sector, as the lattice spacing is
reduced so as to reach the continuum limit. Open boundary condition in the
temporal direction has been proposed and 
investigated \cite{open0,open1,open2} as a (partial) cure to this
problem. Open boundary condition also has been advantageously used in the
investigation of SU(2) lattice gauge theory at weak coupling 
in Ref. \cite{grady}. With open boundary condition, 
one can probe TCDC for even smaller lattice spacings and address questions
related to the scaling etc.

With the adoption of the algebraic
definition of $q(x)$, to suppress unwanted lattice artifacts,
smearing of gauge fields is necessary. In our past study of TCDC
\cite{topocorr}, we have employed HYP smearing \cite{hyp}.
However, the recently proposed Wilson 
flow \cite{wf1,wf2,wf3} makes smearing a 
well-defined mathematical procedure and in addition,
provides a common reference scale to extract physical quantities from
lattice calculations employing different lattice spacings. 
The continuum properties of observables such as TCDC can be studied by
investigating the small flow time behavior.
The scale provided by
Wilson flow has been used in pure Yang-Mills theory to compare
topological susceptibility calculated at different lattice 
spacings \cite{opentopo}. Similar scaling study
for topological susceptibility 
has been performed recently also for dynamical fermions \cite{Bruno}.

With the help of open boundary condition, we have been able to extract
~\cite{Chowdhury:2014kfa} scalar glueball mass down to a 
lattice spacing of 0.0345 fm which was not accessible with periodic
boundary condition. It will be also very interesting to
extract the pseudoscalar glueball mass with the same ensembles
of configurations, from the tail region of the TCDC.
 
In addition to the properties of the correlator of topological charge
density, its localization property itself is also of interest. 
There exists a body of literature on the localization properties of the
low-lying Dirac eigen modes \cite{deForcrand:2006my}, because of their 
connection with the 
topological properties of the QCD vacuum. However, the localization
properties of the topological charge density based on the algebraic
definition involving the field strength $F_{\mu\nu}$ seems to have attracted
little attention except for a preliminary study by the MILC collaboration
\cite{Aubin:2004mp}.  It is interesting to study the Inverse Participation 
Ratio (IPR)
associated with the topological charge density distribution and compare their
behavior under Wilson flow and HYP smearing. A direct visualization of the
effect of Wilson flow on topological charge density distribution of typical
gauge configurations will also shed light on their localization 
properties.                          
\section{Simulation Parameters}
In table \ref{table1}, we present simulation parameters for the HMC 
algorithm used to generate configurations with open and periodic
boundary condition in the temporal direction with unimproved $SU(3)$ Wilson
gauge action. We also give the number of configurations used for measurements
of TCDC and IPR. The `Gap' multiplied by $\tau$ denotes the trajectory interval
between two successive measurements. We have chosen the gaps from our studies
of autocorrelation for TCDC at a fixed value of $r$ (in units of fm), such that
the measurements are statistically independent of each other. The ensemble $P_3$
is an exception, as because of significantly smaller ensemble size in this
case (due to
larger value of $\beta$ and periodic boundary), it was not possible to maintain
large enough gaps between successive measurements. Hence, in this case, errors
are somewhat under-estimated. Lattice spacings for different ensembles are
determined using the results from Refs. \cite{gsw,necco}.
\begin{table}
\begin{center}
\begin{tabular}{|c|c|c|c|c|c|c|l|}
\hline \hline
Lattice & Volume & $\beta$ & $N_{\rm cnfg}$ & Gap & $\tau$&$a[{\rm fm}]$ & $t_{0}/a^2$\\
\hline\hline
{$O_1$}&{$24^3\times48$}&{6.21} &{401} & 392 & {3}& {0.0667(5)} & {6.207(15)}\\
\hline
{$O_2$}&{$32^3\times64$}&{6.42} &{405} & 240 & {4}& {0.0500(4)} & {11.228(31)}\\
\hline
{$O_3$}&{$48^3\times96$}&{6.59} &{458} & 160 & {5} & {0.0402(3)} & {17.630(53)}\\
\hline
{$O_4$}&{$64^3\times128$}&{6.71} &{74} & 160 & {10} & {0.0345(4)} & {24.279(227)}\\
\hline
{$P_1$}&{$24^3\times48$}&{6.21} &{401} & 280 & {3} & {0.0667(5)} & {6.197(15)}\\
\hline
{$P_2$}&{$32^3\times64$}&{6.42} &{401} & 176 & {4} & {0.0500(4)} & {11.270(38)}\\
\hline
{$P_3$}&{$48^3\times96$}&{6.59} &{191} & 64 & {5} & {0.0402(3)} & {18.048(152)}\\
\hline\hline
\end{tabular}
\caption{Simulation parameters for the HMC algorithm. $O$ and $P$ refer to
ensembles with open and periodic boundary condition in
the temporal direction. $N_{\rm cnfg}$ is the total number of measurements
while `Gap' denotes the interval between two successive
measurements in units of $\tau$, the molecular dynamics trajectory length
and $t_{0}/a^2$ is the dimensionless reference Wilson flow time.}
\label{table1}
\end{center}  
\end{table} 

With periodic boundary condition, in order to increase statistics,
source averaging is usually performed for the measurement of TCDC.
However, with open boundary condition, one cannot do so as
translational invariance is lost in the temporal direction.
In order to avoid boundary effects in this case, while calculating
TCDC, the source is kept at the mid-point of the temporal extent and
an averaging over the spatial volume is done.
For a meaningful comparison between the results for open and periodic
lattices, we adapt the same source averaging procedure for the latter
case also.

\section{Numerical Results}
       
\subsection{Topological charge density correlator}
The Topological Charge Density Correlator (TCDC) is given by
\begin{equation}
C(r)=\langle q(x)q(y)\rangle,~~~r=\mid x-y \mid
\end{equation}
where $q(x)$ is the topological charge density.
In order to extract the continuum properties of TCDC from lattice simulations, 
smoothing of gauge fields is necessary if one uses the field theoretic definition
of topological charge density.
Unlike observables such as hadron masses and topological susceptibilites, properties of
TCDC depend on the energy scale at which one is probing the system. Thus when we
compare TCDC at different lattice spacings using smeared gauge configurations,
one has to ensure that the smearing is performed at a given energy scale.

\begin{figure}
\begin{center}   
\includegraphics[width=0.5\textwidth]  
{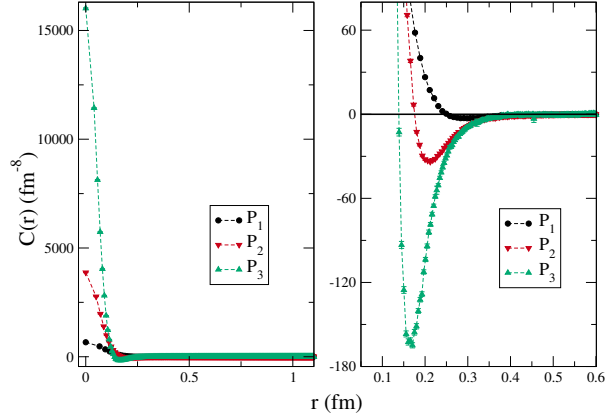}
\caption{ $C(r)$ versus $r$ at 3 HYP smearing steps for 
ensembles $P_1$, $P_2$ and $P_3$. }
\label{comp-diff-beta-pbc-3HYP}
\end{center}
\end{figure}
\begin{figure}[h]
\begin{center}   
\includegraphics[width=0.5\textwidth] 
{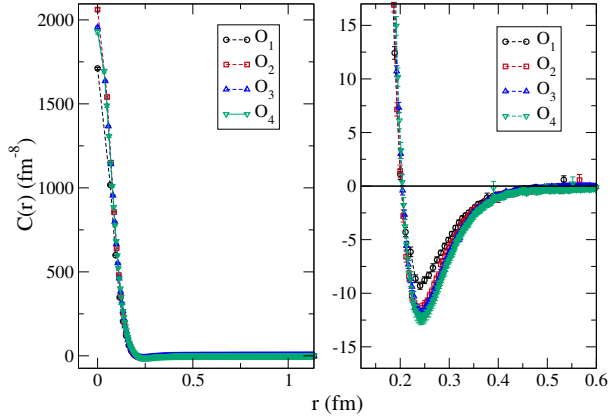}      
\caption{ $C(r)$ versus $r$ at Wilson flow time $\sqrt{8t}=0.14~{\rm fm}$
for ensembles $O_1$, $O_2$, $O_3$ and $O_4$.}                         
\label{cdc-comp-open-diff-beta-014fm}
\end{center}
\end{figure}

In conventional smearing techniques like HYP smearing,
a fixed smearing level does not correspond to a common energy scale to compare 
data generated at different lattice spacings.
In Fig. \ref{comp-diff-beta-pbc-3HYP}, TCDCs at various lattice scales
are compared at the same smearing level 3 (HYP).
From the exhibited behavior one may infer large scaling violations but 
one should keep in mind that a fixed HYP smearing level at different lattice
spacings does not correspond to a common energy scale. 
This is to be contrasted with Wilson flow case, which
facilitates the use of a common energy scale, shown in the
Fig. \ref{cdc-comp-open-diff-beta-014fm}.

Unlike conventional smearing techniques, Wilson flow provides an energy scale
($\frac{1}{\sqrt{8t}}$) 
at which observables can be probed. In order to check possible scaling 
violation in TCDC, one has to choose a particular Wilson flow time for all 
the lattice
spacings investigated. In Fig. \ref{cdc-comp-open-diff-beta-014fm}, we plot
TCDC for ensembles $O_1$, $O_2$, $O_3$ and $O_4$ at the Wilson flow 
time $\sqrt{8t}=0.14~{\rm fm}$
(the rationale for choosing the scale to be $0.14$ fm will be explained later).
Except the data corresponding to the largest lattice spacing, the data 
show universal behavior within our statistical uncertainties.

Recently a comparison of Wilson flow with cooling has been 
performed \cite{Bonati:2014tqa} where a relation between Wilson flow time
and the number of cooling steps has been established. We have investigated 
whether one can phenomenologically 
establish a relation between Wilson flow time and HYP smearing level. In the 
case of TCDC, we found that it is not possible to establish such a relation 
even at a fixed lattice spacing, valid for all $r$. Some approximate relation, of 
course can be found, which however, will vary with lattice spacing. This
issue needs further investigation in the future.

\begin{figure}[h]
\begin{center}   
\includegraphics[width=0.5\textwidth]
{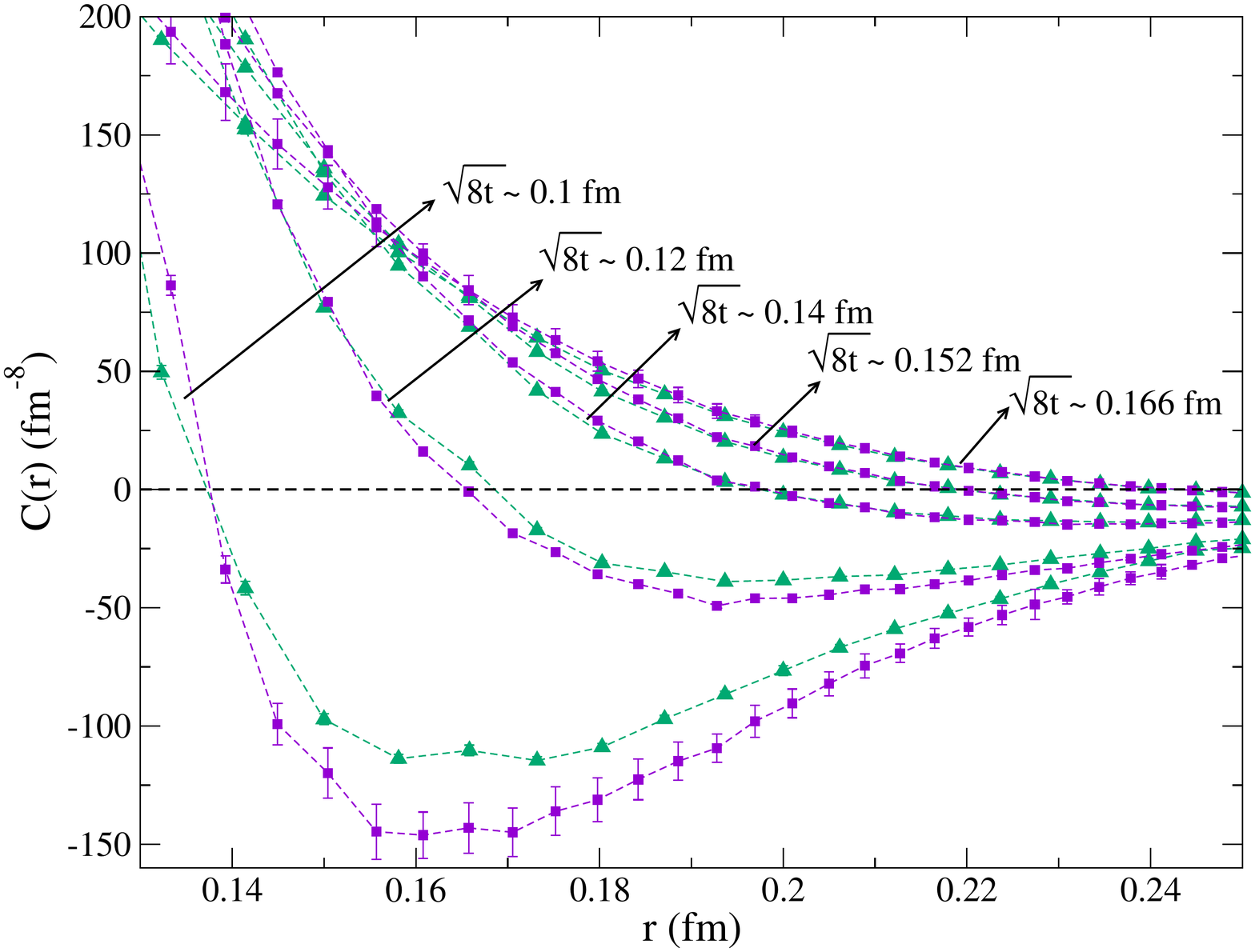}
\caption{Plot of $C(r)$ versus $r$ at various  
values of Wilson flow time $\sqrt{8t}$ at $\beta=6.42$ 
and $\beta=6.59$ for ensembles $P_2$ (filled triangle)
and $P_3$ (filled square) respectively.}   
\label{comp-corr-vs-t-2beta}
\end{center}
\end{figure}
It is expected that the radius of the positive core of $C(r)$ extracted
from lattice data shrinks to zero in the continuum limit. In order to investigate
this phenomena first we need to demonstrate the scaling behavior of $C(r)$
extracted at different lattice spacings, probed at a given Wilson flow
time. Then we need to study the behavior of $C(r)$ as Wilson flow time goes
to zero to extract the continuum behavior.
In Fig. \ref{comp-corr-vs-t-2beta}, we plot $C(r)$ versus $r$ at various  
values of Wilson flow time $\sqrt{8t}$ at $\beta=6.42$ 
and $\beta=6.59$ for ensembles $P_2$ (filled triangle)
and $P_3$ (filled square) respectively. 
At each Wilson flow time probed the data corresponding to
two different lattice scale approximately give the same radius of positive core
of $C(r)$. The agreement is approximate partly because the matching of $\sqrt{8t}$
for two different lattice spacings is approximate. We also observe that
the radius of the positive core of $C(r)$ decreases as Wilson flow time decreases
as expected.

\begin{figure}[h]
\begin{center}
\includegraphics[width=0.5\textwidth]
{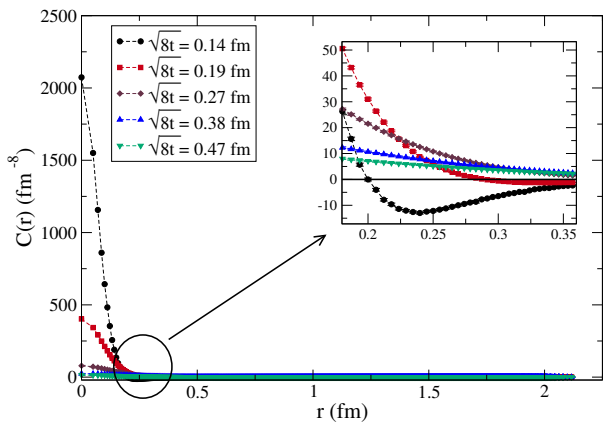}
\caption{Plot of topological charge density correlator 
$C(r)$ versus $r$ at various
value of Wilson flow time $\sqrt{8t}$ at $\beta=6.42$ 
and lattice volume $32^3\times 64$ for ensemble $P_2$.}
\label{cdc-pbc-diff-smlev}
\end{center}
\end{figure}

After investigating the scaling behavior, next, we explore the properties of
$C(r)$ in detail. Without loss of generality we look at the data at a 
particular lattice spacing corresponding to $\beta=6.42$.
In Fig. \ref{cdc-pbc-diff-smlev}, we plot the behavior of $C(r)$ versus $r$
at various Wilson flow times $\sqrt{8t}$ at $\beta=6.42$ 
and lattice volume $32^3\times 64$ for ensemble $P_2$. 
Topological charge density which is constructed from the clover definition of
the field strength tensor further gets extended with Wilson flow time.
As already shown in Fig. \ref{comp-corr-vs-t-2beta}, the size of the positive 
core decreases with decreasing Wilson flow time as presented 
in Fig. \ref{cdc-pbc-diff-smlev}. Note that, further, the heights
of the positive and negative peaks increase with decreasing flow time. 
Since with increasing flow time, the effective size of the charge 
density increases, two charge densities eventually overlap completely
resulting in the disappearance of the negative region of $C(r)$.  

\begin{figure}[h]
\begin{center}   
\includegraphics[width=0.5\textwidth]
{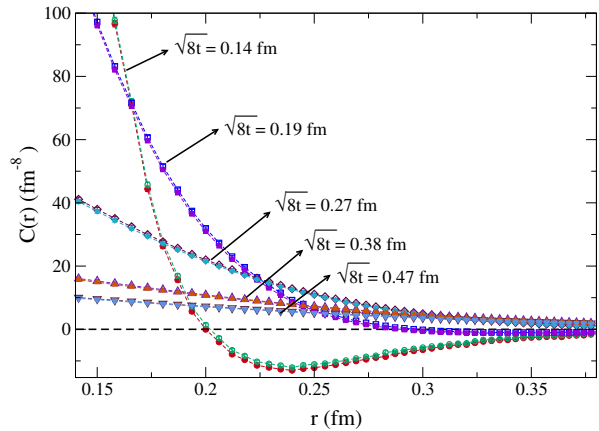}
\caption{Comparison of $C(r)$ versus $r$ at various
value of Wilson flow time $\sqrt{8t}$ at $\beta=6.42$
and lattice volume $32^3\times 64$ for ensembles $P_2$ (filled symbols) and
$O_2$ (open symbols).}
\label{cdc-comp-pbc-open}
\end{center}  
\end{figure}

In Fig. \ref{cdc-comp-pbc-open}, we compare TCDC for the ensembles $P_2$ 
and $O_2$. We note that there is no noticeable
difference between the two TCDC at a given Wilson flow time. Similar trend
has been observed for other lattice spacings as well.

\begin{figure}[h]
\begin{center}   
\includegraphics[width=0.5\textwidth]
{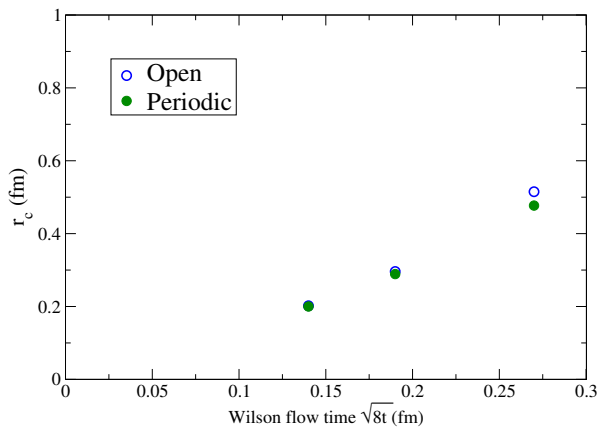}  
\caption{Plot of $r_c$ versus the Wilson flow time $\sqrt{8t}$ for the
ensembles $P_2$ (filled symbols) and $O_2$ (open symbols).}
\label{rc-vs-sm-radius}
\end{center}                       
\end{figure}

In the Fig. \ref{rc-vs-sm-radius}, we plot the the radius of the positive
core ($r_c$) versus the Wilson flow time $\sqrt{8t}$ for the
ensembles $P_2$ (filled symbols) and $O_2$ (open symbols). As noted before,
for larger values of Wilson flow time TCDC is always positive. As
expected, the radius of the positive core $r_c$ diminishes as Wilson flow
time decreases signalling the behavior expected in the continuum. 

\begin{figure}[h]
\begin{center}   
\includegraphics[width=0.5\textwidth]
{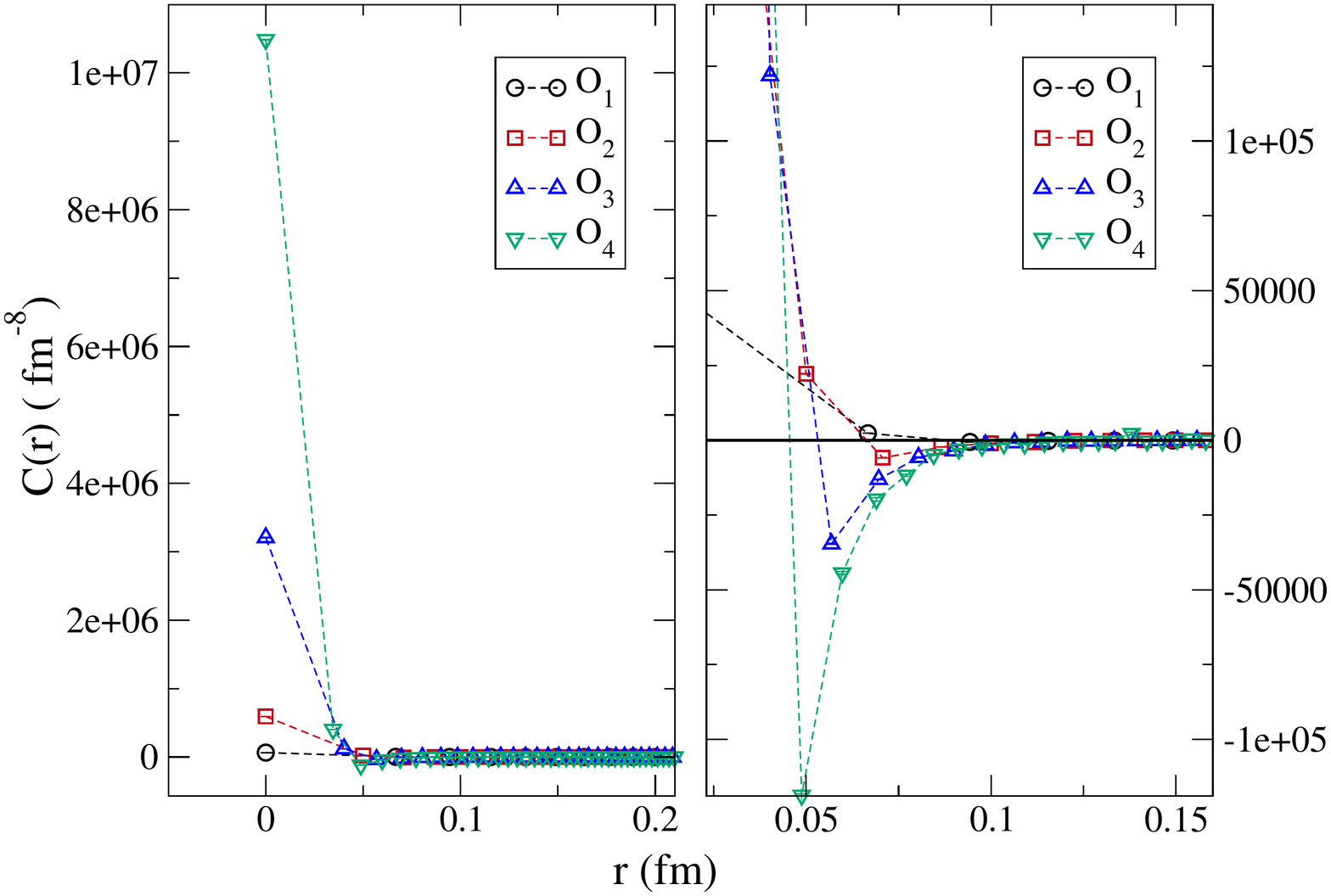}  
\caption{Plot of $C(r)$ versus $r$ for ensembles $O_1$, $O_2$, $O_3$ and $O_4$
without smoothing of gauge fields.}
\label{cdc-comp-open-diff-beta-0fm}
\end{center}                                                                
\end{figure}
In the Fig. \ref{cdc-comp-open-diff-beta-0fm}, we plot the unsmeared TCDC for 
the ensembles $O_1$, $O_2$, $O_3$ and $O_4$. In this case, presence of severe 
lattice artifacts prevents one from extracting any physical observable. For example
one can not extract the topological susceptibility
from such data. Nevertheless we find that the correlator exhibits the 
negativity as expected for the correlator of pseudo scalar operator. Further we 
find that the radius of the positive core
shrinks with lattice spacing as per expectation.

\subsection{Extraction of pseudoscalar glueball mass from TCDC}
Encouraged by the universal behavior exhibited by the TCDC 
for different lattice spacings at a common Wilson flow time, we proceed to
extract the lowest pseudoscalar glueball mass from the tail region of the
TCDC. Due to large vacuum fluctuations present in the correlators of
gluonic observables, the extraction of glueball masses is much more difficult
compared to hadron masses. In QCD, Ref. \cite{Creutz:2010ec} proposed the 
extraction of pseudoscalar flavor singlet meson mass from
topological charge density correlator
to avoid {\em the complexity of separating connected and disconnected quark
diagrams with potentially large statistical fluctuations}. In Yang-Mills 
theory, the extraction of pseudoscalar glueball mass from TCDC
avoids the same complexity. 
Since TCDC has severe singularities and lattice artifacts, smoothing of 
gauge fields is mandatory. Undersmearing of gauge fields leads to persisting
lattice artifacts while oversmearing may wipe out even the negativity character
of the correlator. Thus there is an optimal range of smearing for which one can
reliably extract useful information from the lattice data.
Further, the pseudoscalar glueball mass is expected 
to be much higher than scalar glueball mass. Thus one needs larger statistics and
lower lattice spacings for the extraction of pseudoscalar glueball mass.
Apart from the cost associated with generating configurations, the cost of 
measuring all to all radial correlator also increases rapidly as one goes
to smaller lattice spacings keeping the physical volume constant.

In the past MILC collaboration \cite{bazavov} checked the consistency
of their quenched correlator data using pseudoscalar glueball mass extracted
in Ref. \cite{chen} which used anisotropic lattices, as an input without actually fitting the data.
In this case, a particular level of HYP smearing is used to smooth the 
gauge configurations at different lattice spacings. 


\begin{figure}[h]
\begin{center}   
\includegraphics[width=0.5\textwidth]  
{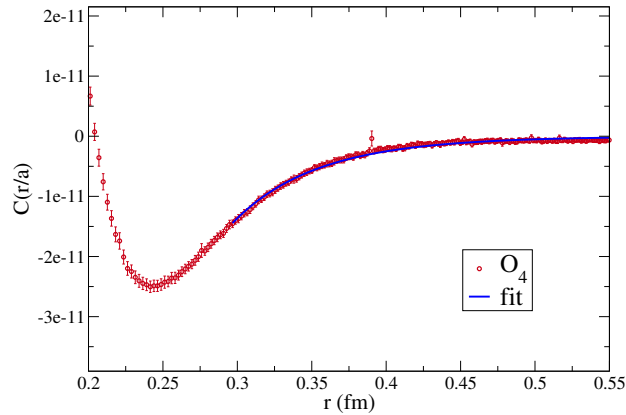}
\caption{ $C(r/a)$ versus $r$ for the ensemble $O_4$ at Wilson flow 
time $\sqrt{8t}$ = 0.14 fm. Also shown is the
the fit used to extract the pseudoscalar glueball mass.}
\label{corr-fit}
\end{center}
\end{figure}

\begin{table} 
\begin{center}
\begin{tabular}{|c|c|c|l|}
\hline \hline
Lattice &$r_{\rm min}$ (fm) &$am$ & m (MeV)\\
\hline\hline
{$O_1$}&0.31&{0.887(39)}&{2624(114)}\\
\hline
{$P_1$}&0.30&{0.831(36)}&{2459(108)}\\
\hline
{$O_2$}&0.29&{0.648(18)}&{2590(78)}\\
\hline
{$P_2$}&0.33&{0.648(25)}&{2560(100)}\\
\hline
{$O_3$}&0.28&{0.535(29)}&{2625(140)}\\
\hline
{$P_3$}&0.27&{0.524(17)}&{2573(81)}\\
\hline
{$O_4$}&0.31&{0.445(11)}&{2545(63)}\\
\hline\hline 
\end{tabular}
\caption{Pseudoscalar glueball mass in lattice and physical units.
The starting point of the fit range is denoted by $r_{\rm min }$.}
\label{table2}
\end{center}
\end{table}

A formula for TCDC is derived in Appendix B of Ref. \cite{sv} for QCD. This 
is in the context of a leading 
term in an effective Lagrangian approach originally proposed by Rosenzweig,
Schechter and Trahern \cite{Rosenzweig:1979ay} and by Di Vecchia and
Veneziano \cite{Di Vecchia:1980ve}. This approach has been further developed, 
being motivated by instanton inspired considerations.
In the analysis of the lattice data,
the following functional
form \cite{sv} of the correlator in the negative region  is
used to extract the pseudoscalar glueball mass: 
\begin{equation}
\langle \phi(x) \phi(y)\rangle = \frac{m}{4 \pi^2 r} K_1(mr)
\end{equation}
where $K_1(z)$ is a modified Bessel function
whose asymptotic form is given by     
\begin{equation}
K_1(z) ~\underset{{\rm large}~z}{\sim} ~
 e^{-z} ~ \sqrt{\frac{\pi}{2z}}~ \left[1 + \frac{3}{8z}  \right ]. 
\label{asym-for}
\end{equation} 

In Fig. \ref{corr-fit}, the TCDC $C(r/a)$ is plotted versus $r$  
for the ensemble $O_4$ at Wilson flow time $\sqrt{8t}$ = 0.14 fm. Also shown 
is the fit in large $r$ region with the formula given in Eq. (\ref {asym-for}) to extract the 
pseudoscalar glueball mass. In the fitting 
procedure the amplitude and the mass are treated as free parameters. We
studied the stability of the fit results (mass and amplitude) with fitting 
range by 
varying both the initial and the final values of $r$ in the tail region of 
TCDC.

\begin{figure}[h]
\begin{minipage}[h]{0.45\linewidth}
\centering
\includegraphics[width=\textwidth]{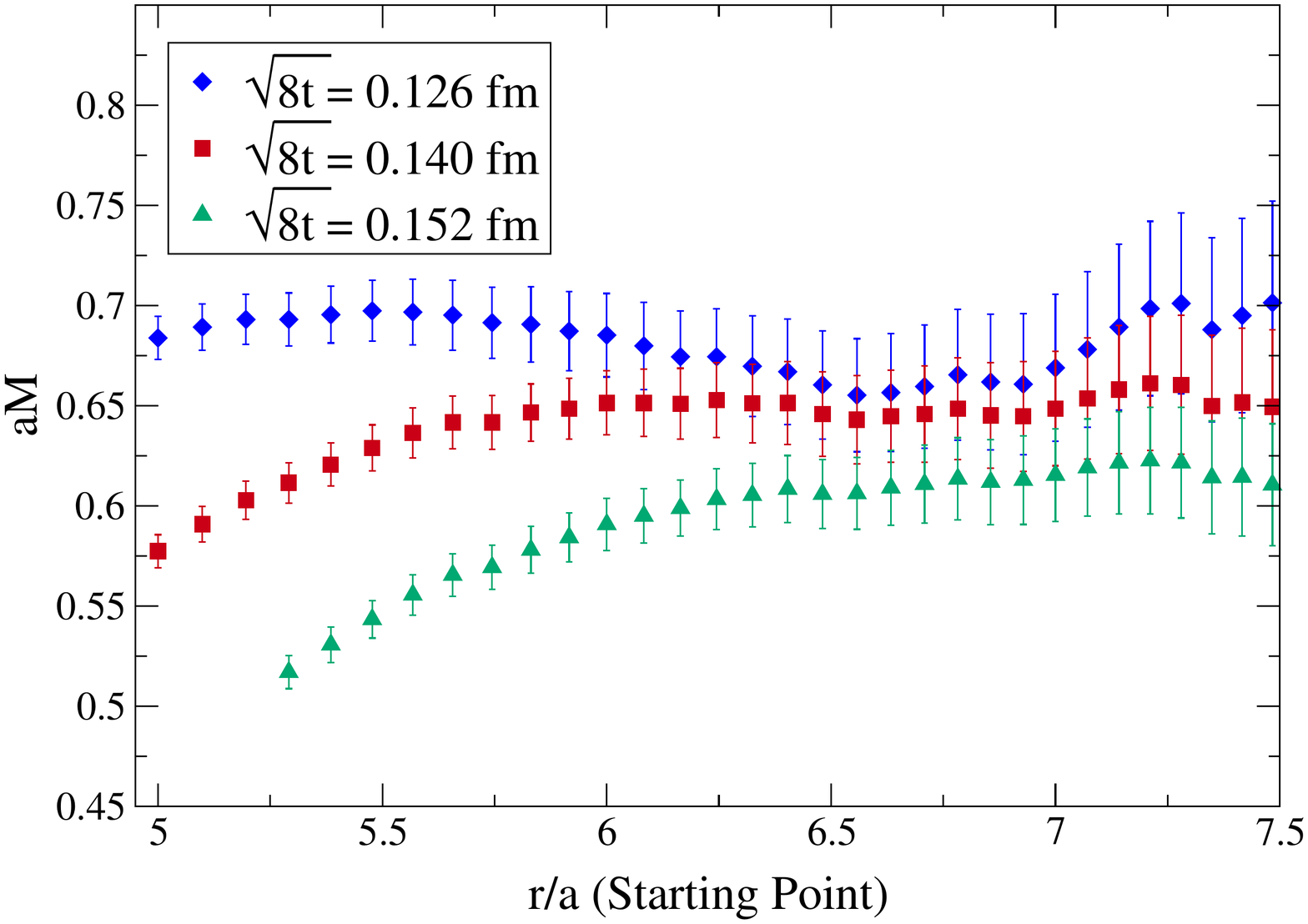}
\end{minipage}
\begin{minipage}[h]{0.45\linewidth}
\centering
\includegraphics[width=\textwidth]{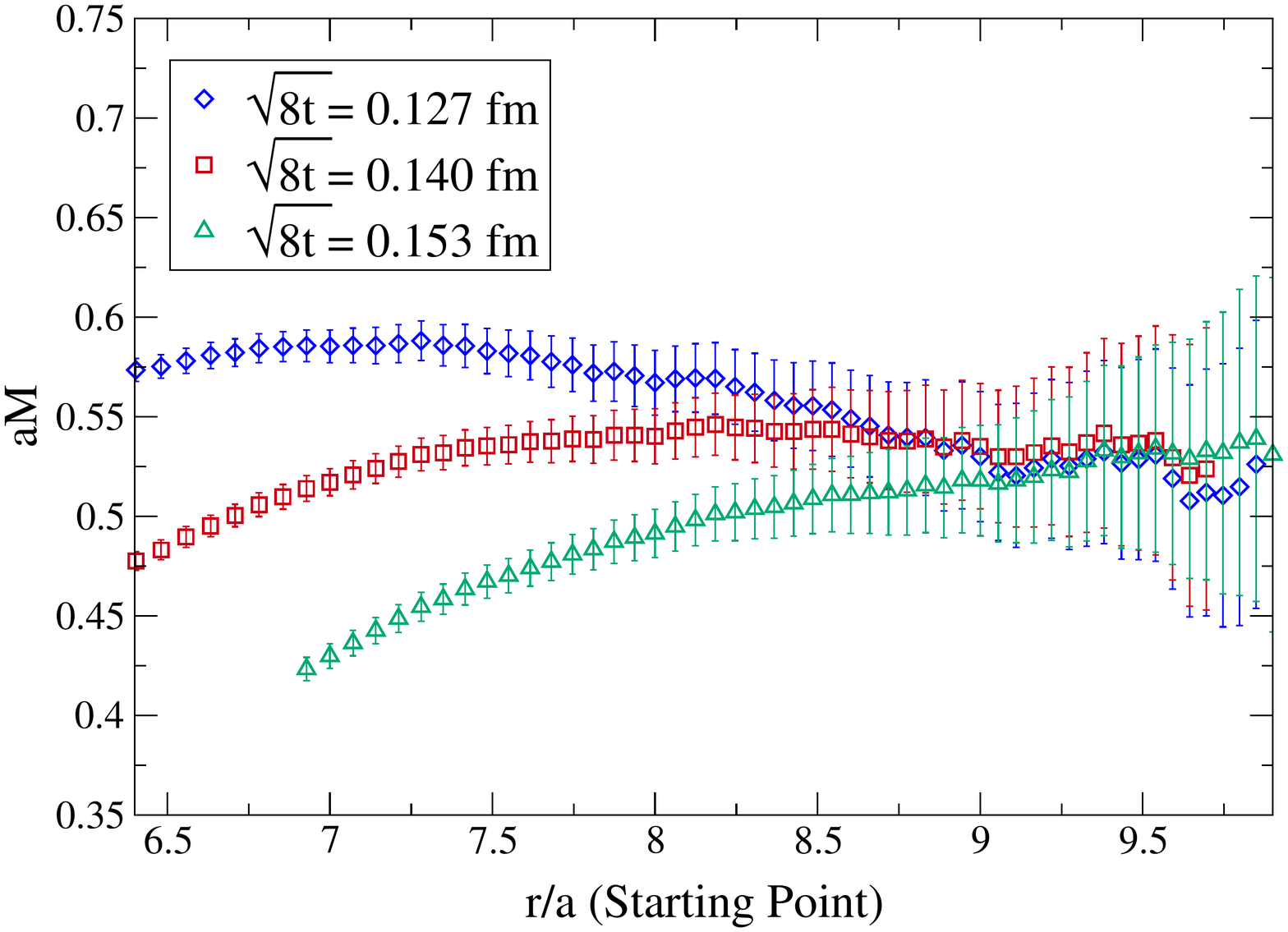}
\end{minipage}
\caption{Sensitivity, to starting point of fit range for fixed end point, of
pseudoscalar glueball mass extracted using the asymptotic formula
for ensemble $P_2$ (left) and ensemble $O_3$ (right).} 
\label{sens}
\end{figure}

We have studied the behavior of the extracted pseudoscalar glueball mass under 
different Wilson
flow times. 
In Fig. \ref{sens}, we show the   
sensitivity of the extracted mass to starting 
point of fit range for a chosen end point,  for the ensembles $P_2$ (left)
and $O_3$ (right). The end point is then varied over a very wide range and
examined the stability of the fit for a given starting point. 
We found that the extracted mass is 
independent of the end point once the error bar of the tail of the
correlator touches the value zero.  
The extracted mass is apparently stable at and around the scale $0.14$ fm as 
exhibited in the overlap of plateau regions of data for three different Wilson flow times.
The overlap between the plateau regions is better for the ensemble $O_3$
compared to the ensemble $P_2$. This behavior is expected as one approaches
the continuum. The extracted mass values for different ensembles are given in 
table \ref{table2}. Since the extracted mass is independent of the end
point, we do not quote a specific value in the table. 
As ensemble sizes for $P_3$ and $O_4$ are comparatively
small, results for these two ensembles may have some bias leading
to possible underestimation of errors as well.


\begin{figure}
\begin{center}   
\includegraphics[width=0.5\textwidth]  
{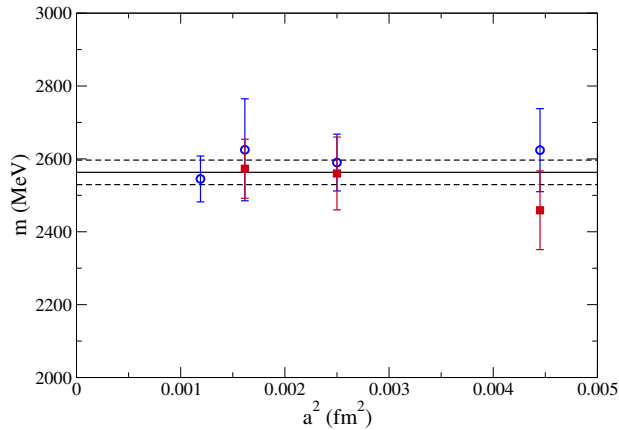}
\caption{Plot of lowest pseudoscalar glueball mass versus $a^2$ for both
open (open symbols) and periodic (filled symbols) boundary conditions at Wilson flow time $\sqrt{8t}$ = 0.14 fm.
Also shown is the fit to 
the data.}
\label{psgbmass}
\end{center}
\end{figure}

In order to extract the pseudoscalar glueball mass in the continuum, in
Fig. \ref{psgbmass} we plot the mass in MeV versus $a^2$ in fm$^{2}$ for both
the boundary conditions for all the lattice spacings explored in 
this work.
As expected from the universal scaling behavior 
exhibited by $C(r)$ in the asymptotic
region (see Fig. \ref{cdc-comp-open-diff-beta-014fm}) within the statistical 
error, the data for mass does not show any deviation from scaling. Hence we 
fit a constant to the data as shown in the figure and thus extract the continuum
value of the pseudoscalar glueball mass as 2563 (34) MeV. This
value compares very well with the value 2560 (35) MeV quoted in Ref. \cite{chen},
which is extracted from the decay of the temporal pseudoscalar correlator on an 
anisotropic lattice.

\subsection{ Localization properties of topological charge density}
\begin{figure}[h]
\begin{center}
\includegraphics[width=0.5\textwidth]
{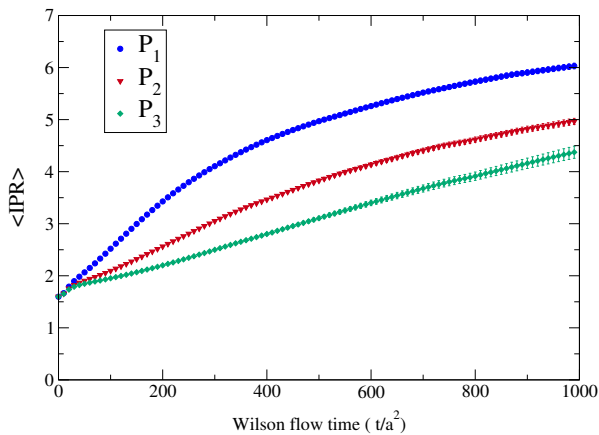}
\caption{Plot of configuration average of Inverse Participation Ratio 
($\langle {\rm IPR}\rangle$) versus Wilson flow time 
($t \slash a^2$) 
for the ensembles $P_1$, $P_2$ and $P_3$. }
\label{ipr-wf-pbc}
\end{center}
\end{figure}

It is interesting to study the localization property of the topological charge density
$q(x)$ and its behavior under Wilson flow. Since Wilson flow time provides an energy
scale to probe the system, it becomes possible to extract the continuum
behavior by studying the small Wilson flow time behavior of the observables.

As a measure of the localization property of $q(x)$, 
one can use the Inverse Participation Ration (IPR)  
defined as~\cite{deForcrand:2006my,Aubin:2004mp}
\begin{equation}
{\rm IPR} = V \frac{\sum_x \mid q(x)\mid^4}{(\sum_x \mid q(x)\mid^2)^2}
\end{equation}
where $V$ is the four dimensional lattice volume. If $q(x)$ is completely local, 
for example $q(x)=\delta (x_0)$, $IPR=V$ and if $q(x)$ is completely delocalized,
$q(x)= c$, a constant then  $IPR=1$. If $q(x)$ is localized on a fraction of
sites $f$ then  $IPR=\frac{1}{f}$~\cite{Aubin:2004mp}. Thus measurement of
IPR provides information about the localization properties of $q(x)$.
For an excellent discussion on IPR \cite{bd} which was originally introduced
in the context of condensed matter
systems, see Ref. \cite{wegner}.

To investigate the effect of Wilson flow on the localization property of 
topological charge density, we plot the configuration average of IPR versus
Wilson flow time ($t \slash a^2$) in Fig. \ref{ipr-wf-pbc}
for the ensembles $P_1$, $P_2$ and $P_3$. We note that $\langle IPR \rangle$
monotonously decreases with decreasing flow time ($t \slash a^2$) indicating
reduced localization of $q(x)$. Also, {\em throughout} the range of $t \slash a^2$,
$\langle IPR \rangle$ appears to decrease with decreasing lattice spacing.

\begin{figure}[h]
\begin{center}
\includegraphics[width=0.5\textwidth]
{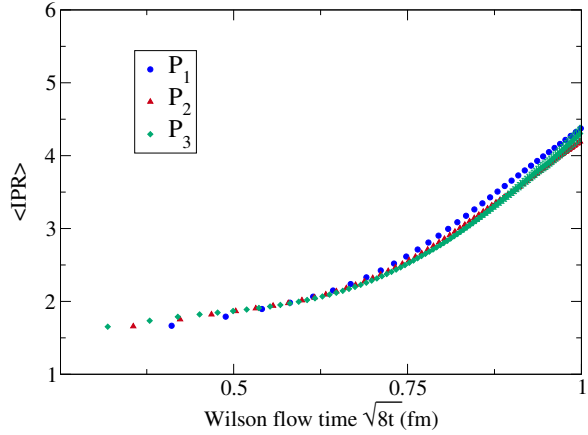}
\caption{$\langle IPR \rangle$ versus $ \sqrt{8t} $ for 
ensembles $P_1$, $P_2$ and $P_3$.} 
\label{ipr-wf-scaled-pbc}
\end{center}
\end{figure}

Since Wilson flow provides a scale (independent of the lattice spacing) 
to probe the observables, it is more interesting to study the variation of
$\langle {\rm IPR} \rangle$ with respect to $\sqrt{8t}$ for ensembles at
different lattice spacings. In Fig. \ref{ipr-wf-scaled-pbc} we plot
$\langle {\rm IPR} \rangle$ versus $ t \slash r_0^2$ for ensembles $P_1$, $P_2$ 
and $P_3$. We find that, remarkably, unlike the behavior shown in figure
\ref{ipr-wf-pbc},
the average IPR's for different ensembles are now very close to each other 
and the average IPR's for the ensembles corresponding to the smaller two 
lattice spacings agree with each other within our statistical accuracy.
The data for ensembles corresponding to the largest lattice spacing ($\beta=6.21$)
exhibits some mild scaling violation as already noted in the case of TCDC in
Fig. \ref{cdc-comp-open-diff-beta-014fm}. We note that, interestingly,
IPR is small for small Wilson flow time and monotonously increases
as Wilson flow time increases. This indicates that when probed at short
distances $q(x)$ is very much delocalized. On the other hand when probed
at long distances $q(x)$ appears more localized.
   
\begin{figure}[h]
\begin{center}
\includegraphics[width=0.5\textwidth]
{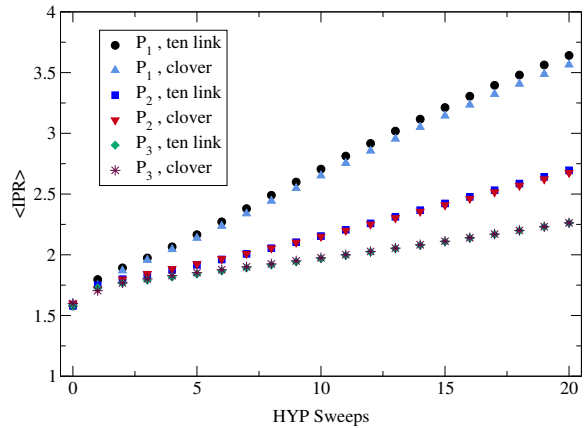}
\caption{Plot of $\langle {\rm IPR} \rangle$ versus HYP sweeps using two 
definitions
of the topological charge density (clover and ten link) for
ensembles $P_1$, $P_2$ and $P_3$.}
\label{ipr-hyp-pbc}
\end{center}
\end{figure}

Among the available algebraic definitions for $q(x)$, 
one is based on the clover expression for the field strength $F_{\mu\nu}$
which is the simplest. 
Another is 
the more sophisticated ten-link definition developed for $SU(2)$ by 
DeGrand, Hasenfratz and Kovacs \cite{degrand}, modified for $SU(3)$ by 
Hasenfratz and Neiter \cite{hasenfratz1}. Also there are studies employing
tree level improvement over the clover definition of $q(x)$ (as well as the
action for cooling process) in case of $SU(2)$ gauge theory \cite{Philipe}.
In Fig. \ref{ipr-hyp-pbc}, we plot $\langle {\rm IPR} \rangle$ versus HYP
sweeps using two definition
of topological charge density (clover and ten link) for
ensembles $P_1$, $P_2$ and $P_3$. It is expected that different lattice
discretizations of the topological charge density yield the same physical
observable as one approaches the continuum limit. We note that the  
$\langle {\rm IPR} \rangle$ for the two definitions of topological charge
density move closer to each other with decreasing lattice spacing, as expected.

\begin{figure}[t]
\begin{minipage}[h]{0.45\linewidth}
\centering
\includegraphics[width=\textwidth]{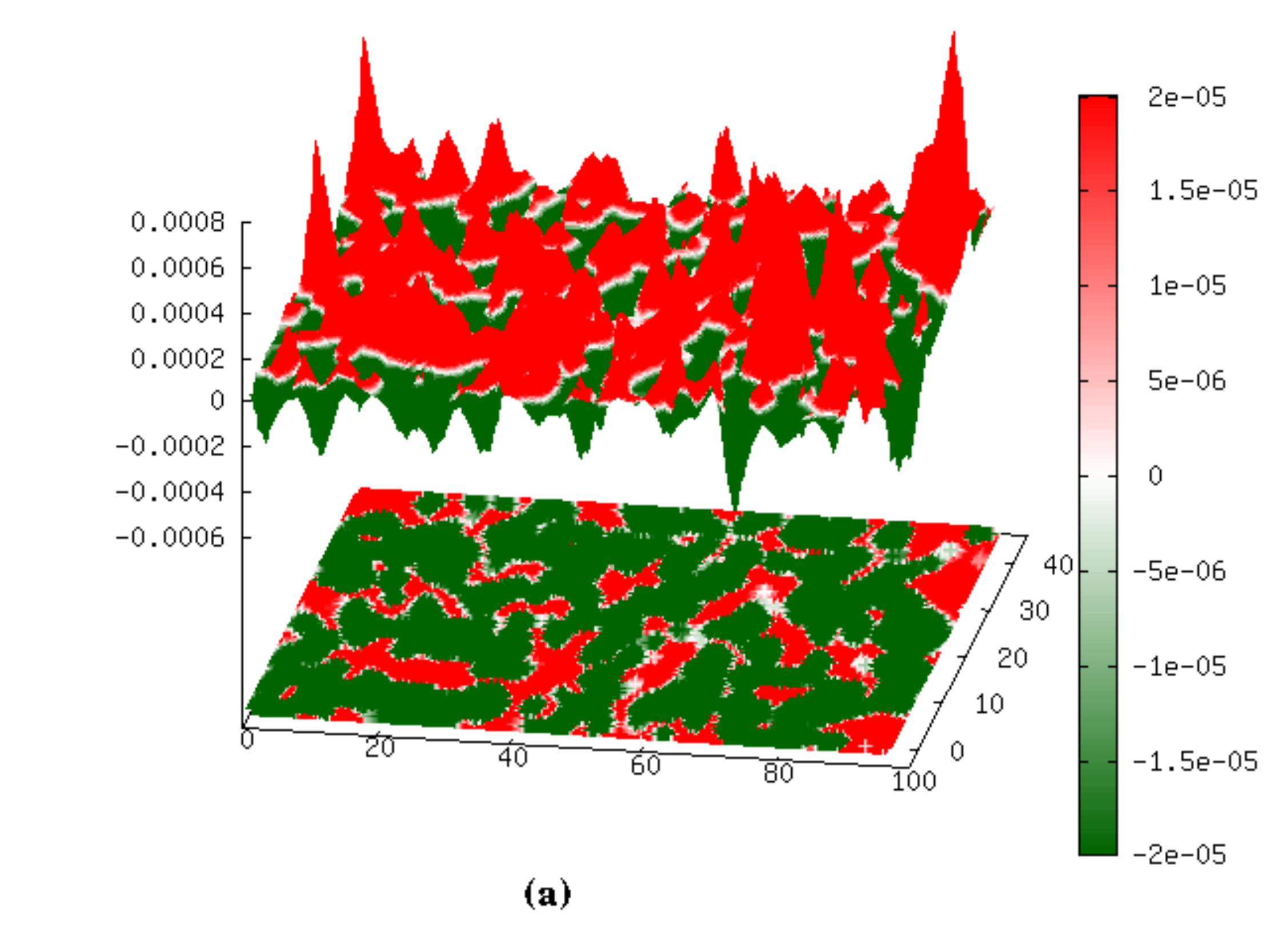}
\end{minipage}
\begin{minipage}[h]{0.45\linewidth}
\centering
\includegraphics[width=\textwidth]{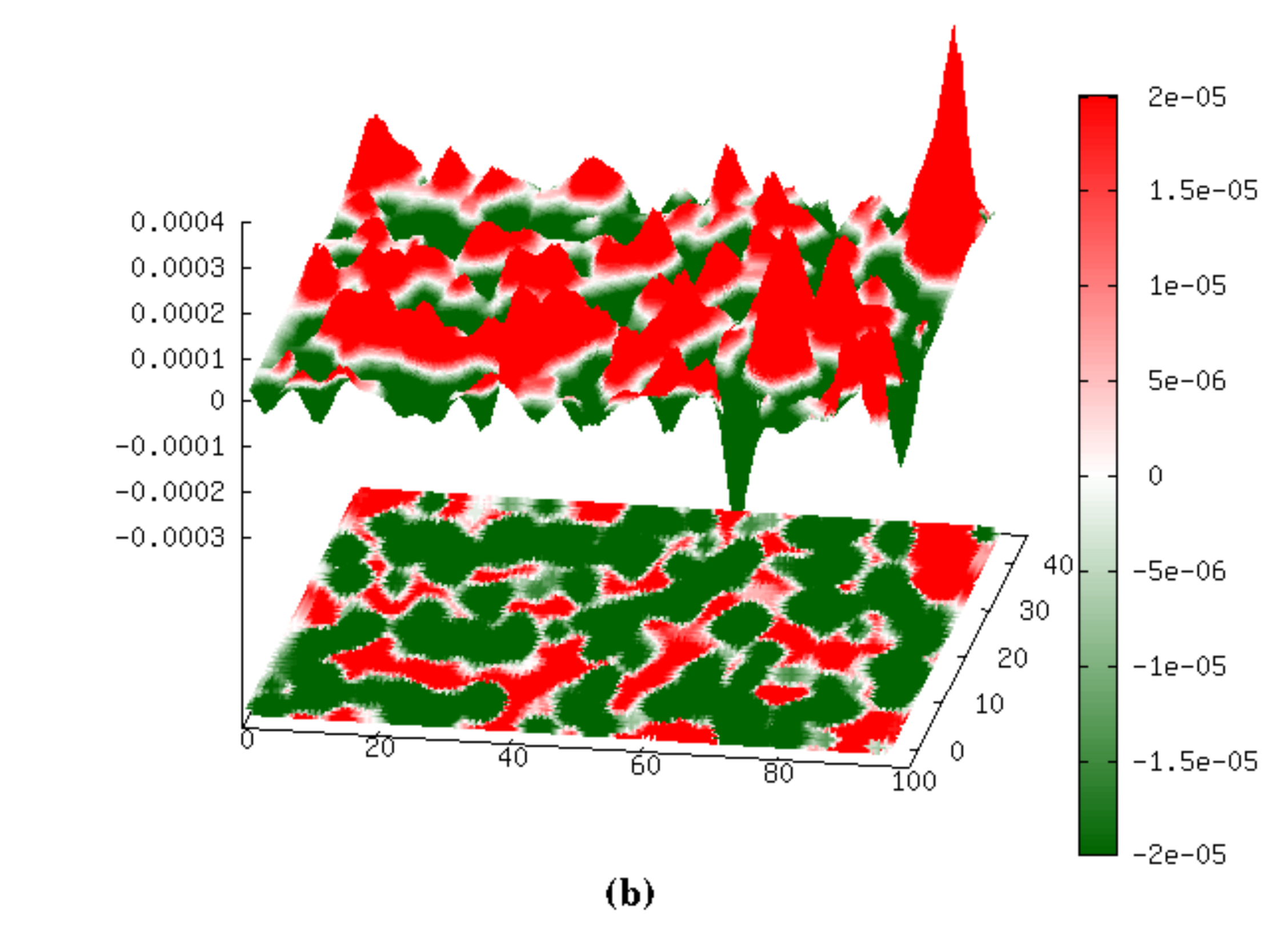}
\end{minipage}
\begin{minipage}[h]{0.45\linewidth}
\centering
\includegraphics[width=\textwidth]{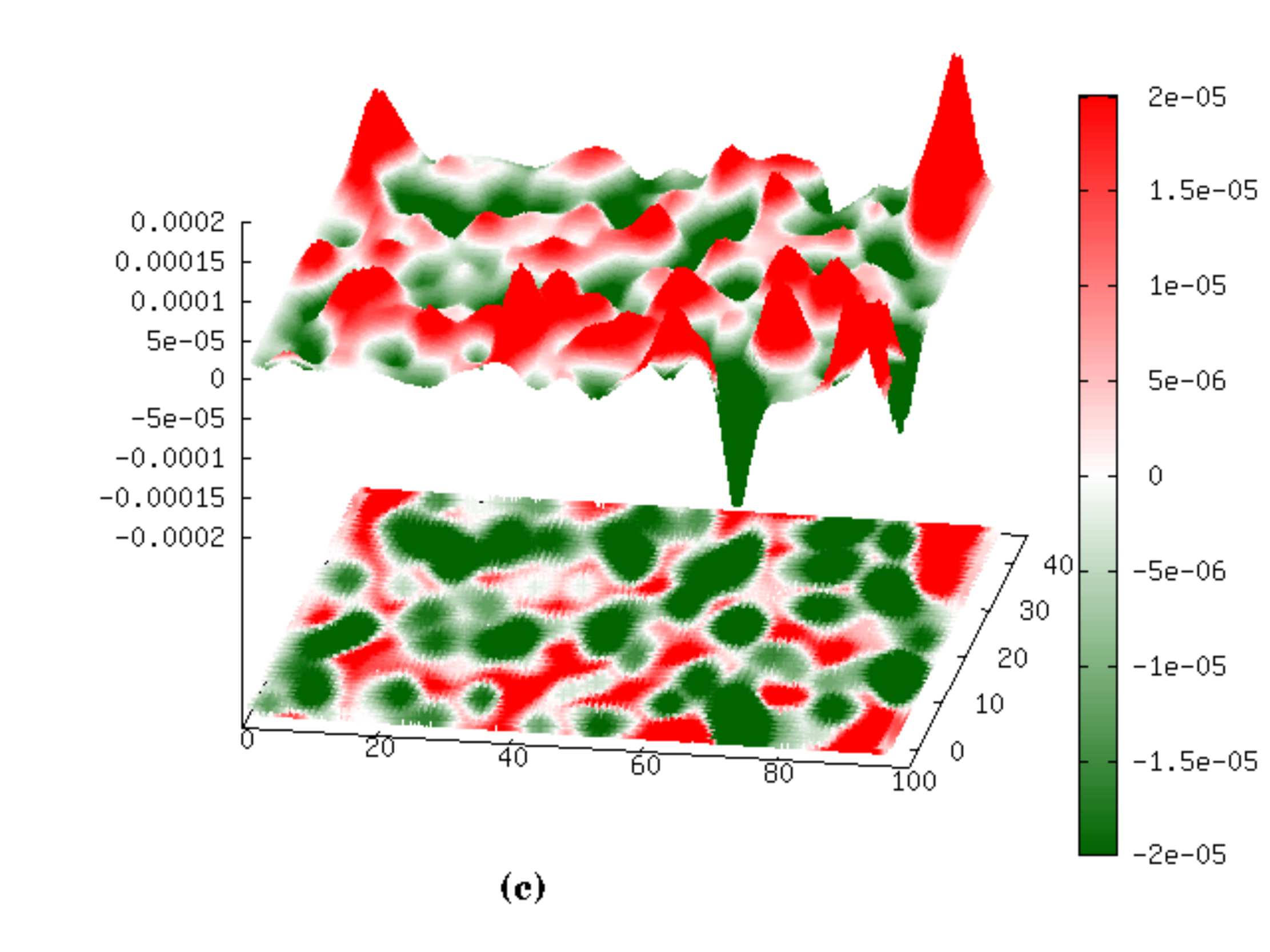}
\end{minipage}
\begin{minipage}[h]{0.45\linewidth}
\centering
\includegraphics[width=\textwidth]{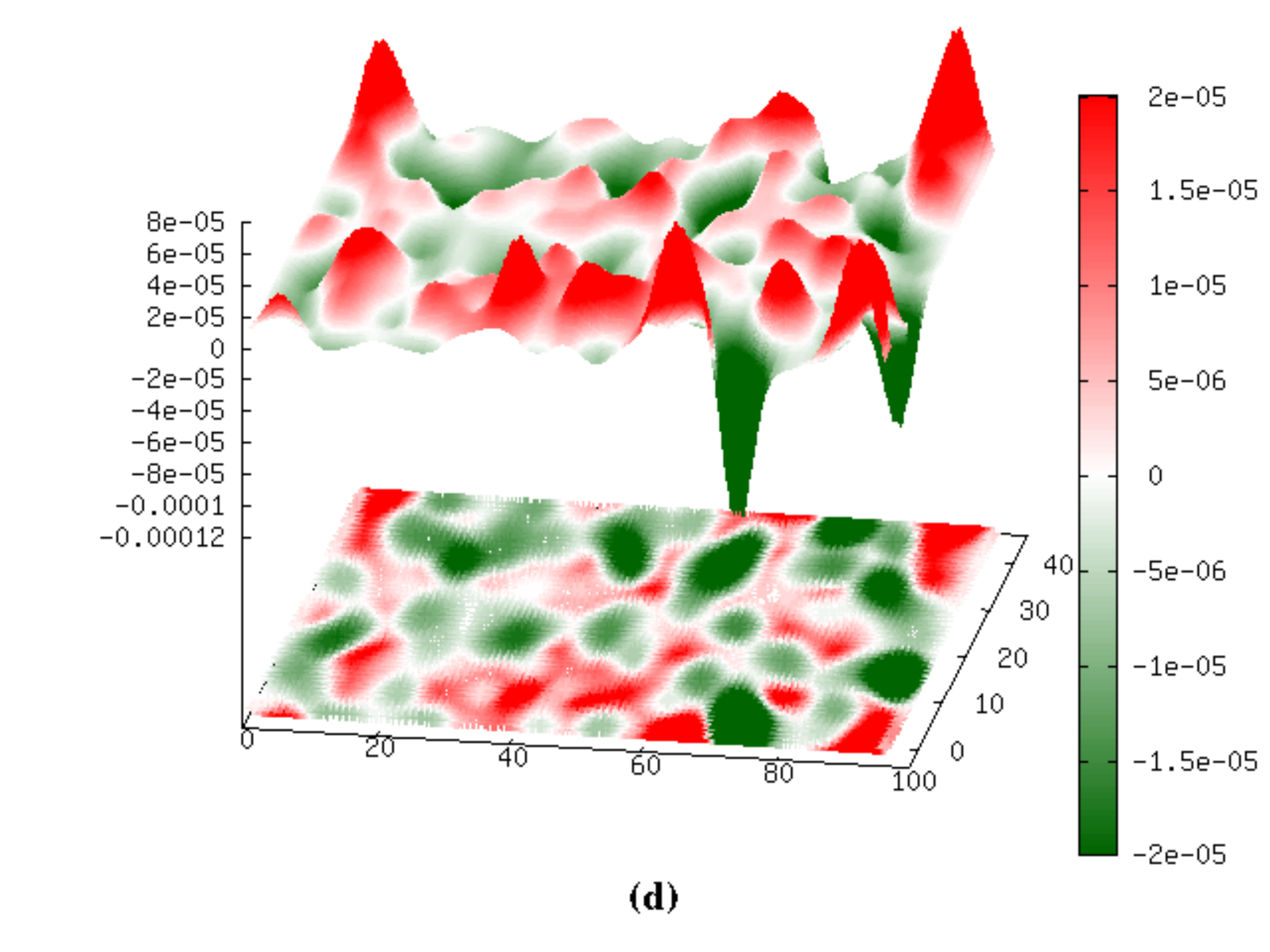}
\end{minipage}
\begin{minipage}[h]{0.45\linewidth}
\centering
\includegraphics[width=\textwidth]{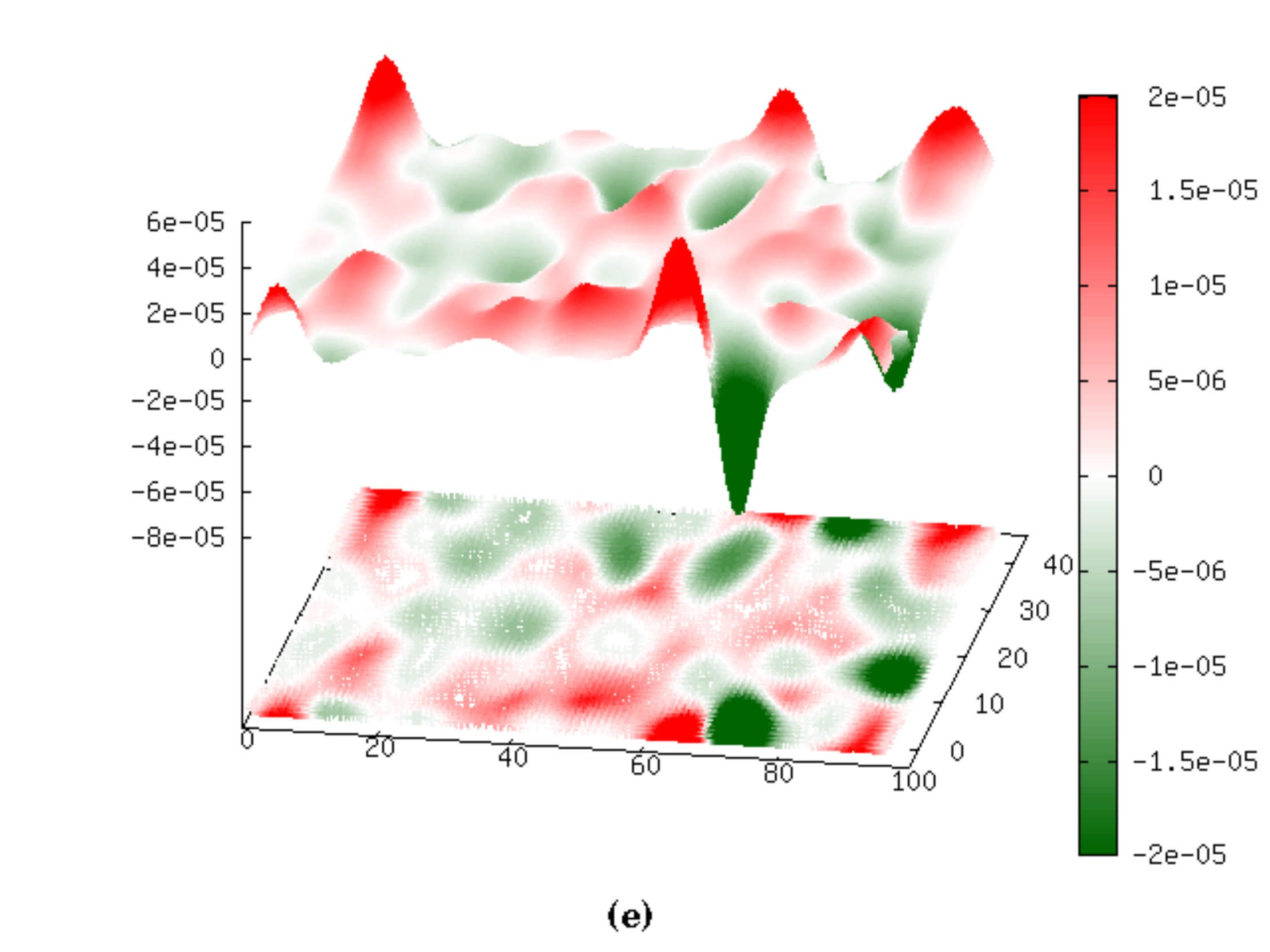}
\end{minipage}
\begin{minipage}[h]{0.45\linewidth}
\centering
\includegraphics[width=\textwidth]{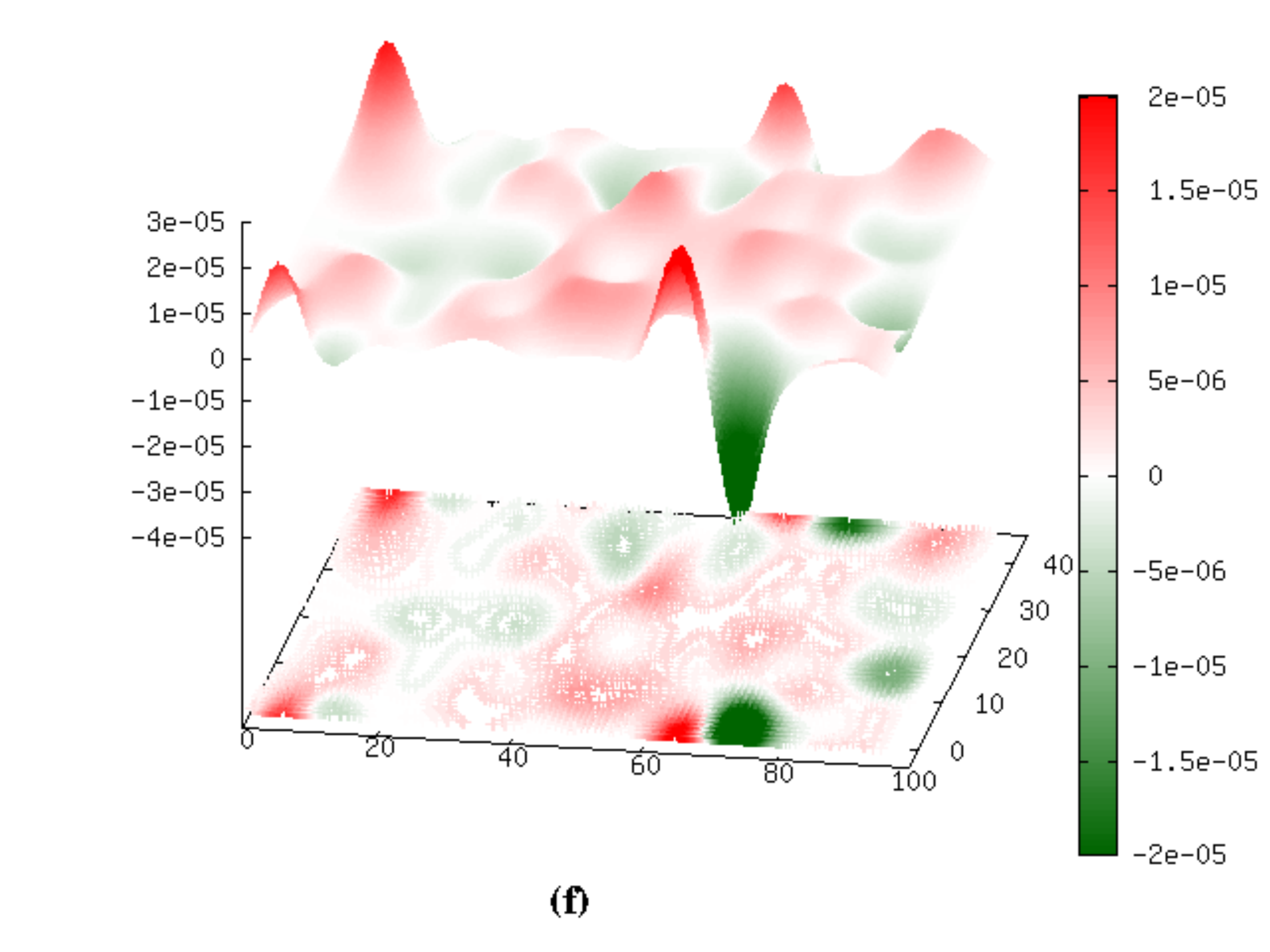}
\end{minipage}
\caption{The behavior of the topological charge density distribution 
$q(x)$ under Wilson flow, as a function of $x_0$ and $x_1$ at $x_2=x_3=24$ 
for a typical configuration belonging to the ensemble $O_3$. 
The plots (a) to (f) correspond to the flow times $\sqrt{8t}$ = 0.14, 0.19, 
0.25, 0.3, 0.38 and 0.47 fm respectively.} 
\label{charge-profile}
\end{figure}

We also note that $\langle {\rm IPR}
\rangle$ decreases with decreasing lattice spacing at a given flow time.
The behavior we have observed appears compatible with that exhibited by
the data of MILC collaboration \cite{Aubin:2004mp} at their three smaller
lattice spacings. Note that the largest lattice spacing explored in our
work is smaller than the smallest lattice spacing studied in 
Ref. \cite{Aubin:2004mp} which, however, has employed an improved lattice
action.

In order to gain a better understanding of the behavior of both 
the charge density correlator $C(r)$ and the inverse participation ratio
(IPR) we plot, in Fig. \ref{charge-profile}, the behavior of the topological
charge density distribution $q(x)$ under Wilson flow, as a function of $x_0$ 
and $x_1$ at $x_2=x_3=24$ for a typical configuration belonging to the 
ensemble $O_3$. The plots (a) to (f) correspond to the flow times $\sqrt{8t}$ 
= 0.14, 0.19, 0.25, 0.3, 0.38 and 0.47 fm respectively. 
In the continuum, it is expected that TCDC possess a positive core and 
a negative peak adjacent to each other and close to origin.
This continuum behavior can be probed through the small values
of Wilson flow time. At relatively small values of the Wilson flow time, 
it is seen that $q(x)$ possesses regions of 
both positive and negative charge densities of relatively large magnitudes
lying next to each other. This provides a qualitative explanation~\cite{alt} of the
positive core and the adjoining negative peak observed in $C(r)$. As the
Wilson flow time decreases, the proximity of the regions of positive and
negative charge densities of large magnitudes increases, and the charge
density appears to be more delocalized. This results in increased participation for $q(x)$ 
which in turn, explains the decrease of IPR with decreasing Wilson flow time
as discussed previously.
     
In the past, there have been studies (see, for example, 
Refs. \cite{ih, Ilgenfritz:2007xu,Aubin:2004mp}
of the fractal
dimension of the topological charge density, which have used the
Ginsparg-Wilson definition of the topological charge density and improved
actions, but at considerably large lattice spacings compared to present
standards. Our study of the IPR, on 
the other hand, has utilized the Wilson action at much smaller lattice
spacings and the field strength definition of the topological charge density
together with the Wilson flow to smooth the gauge configurations. It will be
very interesting to extend the previous studies of the fractal dimension    
to much smaller lattice spacings together with the use of Wilson flow to    
study the fractal diemnsion as a function of the flow time, to avoid any    
possible contamination of lattice artifacts so that one can reach definite  
conclusions. Our current study nevertheless seems to support the    
notion that 
{\em the structure of the vacuum of Yang-Mills theory depends on the scale
at which it is probed} \cite{deForcrand:2006my}.

\section{Conclusions}
In SU(3) Lattice Yang-Mills theory, we have investigated TCDC and IPR of
topological charge density ($q(x)$) for a range of relatively small
lattice spacings with the view of studying the continuum properties.
As expected we have not found any noticeable difference between periodic
and the recently proposed open boundary conditions. However, open boundary
condition has enabled us to compute observables at a smaller lattice
spacing because of the absence of the {\em trapping problem}.
Recently proposed Wilson flow, in contrast to the smearing techniques
proposed previously, provides a common energy scale to probe the
system simulated for variety of lattice spacings.
In contrast to a fixed HYP smearing level, 
by choosing a particular Wilson flow time $t$ for all the
lattice spacings investigated, we find that, except the data corresponding to 
the largest lattice spacing, the TCDC data show universal behavior within our 
statistical uncertainties. 
The continuum properties of TCDC are inferred by the studying
the small flow time behavior.
The pseudoscalar glueball mass obtained from the tail region
of TCDC does not exhibit any noticeable scaling violation and the
extracted value in the continuum, 2563 (34) MeV agrees well with
the value extracted previously in the literature with anisotropic lattices.
We found that the configuration average of the inverse participation ratio for
topological charge density, calculated at a given level of smearing
(for the Wilson flow it is $t/a^2$) decreases with the lattice spacing. However 
when plotted versus common scale $\sqrt{8t}$,
it seems to be independent of lattice spacing.
A detailed study of $q(x)$ under Wilson flow time revealed that as
Wilson flow time decreases, the proximity of the regions of positive and
negative charge densities of large magnitudes increases, and the charge
density appears to be more delocalized resulting in the observed decrease of IPR
with decreasing Wilson flow time.

\begin{acknowledgments}
To carry out all the numerical calculations reported in this work, Cray XT5 
and Cray XE6 systems supported by the 11th-12th Five Year
Plan Projects of the Theory Division, SINP under the Department of 
Atomic Energy, Govt. of India, are used.
We thank Richard Chang for the prompt maintenance of the systems
and the help in data management.
For some useful discussions, we also thank Pushan Majumdar and Santanu Mondal. 
We are indebted to Philippe de Forcrand for helpful comments on the earlier
version of this manuscript.
This work was in part based on the publicly available lattice gauge theory 
code {\tt openQCD} \cite{openqcd} and that of MILC collaboration \cite{milc} .
\end{acknowledgments}


\end{document}